\documentclass[11pt,a4paper]{emulateapj}

\usepackage{epsfig}
\usepackage{amsmath}

\def\apj{ApJ}
\def\apjl{ApJL}
\def\mnras{MNRAS}
\def\pasp{PASP}

\def\araa{ARAA}
\def\aap{A\&A}
\def\aj{AJ}
\def\apjs{ApJS}

\def\nat{Nature}

\def\gs{\mathrel{\raise0.35ex\hbox{$\scriptstyle >$}\kern-0.6em\lower0.40ex\hbox{{$\scriptstyle \sim$}}}} 
\def\ls{\mathrel{\raise0.35ex\hbox{$\scriptstyle <$}\kern-0.6em\lower0.40ex\hbox{{$\scriptstyle \sim$}}}}

\def\Wm2{\,\hbox{W}\,\hbox{m}^{-2}} 
\def\gsim{\mathrel{\raise0.35ex\hbox{$\scriptstyle >$}\kern-0.6em\lower0.40ex\hbox{{$\scriptstyle \sim$}}}} 
\def\lsim{\mathrel{\raise0.35ex\hbox{$\scriptstyle <$}\kern-0.6em\lower0.40ex\hbox{{$\scriptstyle \sim$}}}} 
\def\ltsima{$\; \buildrel < \over \sim \;$} 
\def\simlt{\lower.5ex\hbox{\ltsima}} 
\def\gtsima{$\; \buildrel > \over \sim \;$} 
\def\simgt{\lower.5ex\hbox{\gtsima}}

\lefthead{Swinbank et al.}  \righthead{The Star-Forming ISM at $z$\,=\,0.84--2.23 from HiZELS}

\begin{document}

\title {The Properties of the Star-Forming Interstellar Medium at
  \lowercase{$z$}\,=\,0.8--2.2 from HiZELS: Star-Formation and Clump
  Scaling Laws in Gas Rich, Turbulent Disks}

\author{
A.\,M.\ Swinbank,\altaffilmark{1}
Ian Smail,\altaffilmark{1}
D.\ Sobral,\altaffilmark{2,3}
T.\, Theuns,\altaffilmark{1,4}
P.\,N.\ Best,\altaffilmark{3}
\& J.\,E.\ Geach,\altaffilmark{5}}
\setcounter{footnote}{0}
\altaffiltext{1}{Institute for Computational Cosmology, Department of Physics, Durham University, South Road, Durham DH1 3LE, UK; email: a.m.swinbank@dur.ac.uk}
\altaffiltext{2}{Leiden Observatory, Leiden University, PO Box 9513, 2300 RA Leiden, the Netherlands}
\altaffiltext{3}{SUPA, Institute for Astronomy, University of Edinburgh, Edinburgh, EH19 3HJ, UK}
\altaffiltext{4}{University of Antwerp, Campus Groenenborger, Groenenborgerlaan 171, B-2020 Antwerp, Belgium}
\altaffiltext{5}{Department of Physics, McGill University, Ernest Rutherford Building, 3600 Rue University, Montreal, Quebec H3A 2T8, Canada}

\begin{abstract}
  We present adaptive optics assisted integral field spectroscopy of
  nine H$\alpha$-selected galaxies at $z$\,=\,0.84--2.23 drawn from the
  HiZELS narrow-band survey.  Our observations map the kinematics of
  these star-forming galaxies on $\sim$\,kpc-scales.  We demonstrate
  that within the ISM of these galaxies, the velocity dispersion of the
  star-forming gas ($\sigma$) follows a scaling relation
  $\sigma\propto\Sigma_{\rm SFR}^{1/n}$\,+$\,constant$ (where
  $\Sigma_{\rm SFR}$ is the star formation surface density and the
  constant includes the stellar surface density).  Assuming the disks
  are marginally stable (Toomre $Q$\,=\,1), this follows from the
  Kennicutt-Schmidt relation ($\Sigma_{\rm SFR}$\,=\,$A\Sigma_{\rm
    gas}^n$), and we derive best fit parameters of
  $n$\,=\,1.34\,$\pm$\,0.15 and
  $A$\,=\,3.4$_{-1.6}^{+2.5}\times$\,10$^{-4}$\,M$_{\odot}$\,yr$^{-1}$\,kpc$^{-2}$,
  consistent with the local relation, and implying cold molecular gas
  masses of M$_{\rm gas}$\,=\,10$^{9-10}$\,M$_{\odot}$ and molecular
  gas fractions M$_{\rm gas}$\,/\,(M$_{\rm
    gas}$\,+\,M$_{\star}$)\,=\,0.3\,$\pm$\,0.1, with a range of
  10\,--\,75\%.  We also identify eleven $\sim$\,kpc-scale star-forming
  regions (clumps) within our sample and show that their sizes are
  comparable to the wavelength of the fastest growing mode.
  The luminosities and velocity dispersions of these clumps follow the
  same scaling relations as local H{\sc ii} regions, although their
  star formation densities are a factor $\sim$\,15\,$\pm$\,5\,$\times$
  higher than typically found locally.  We discuss how the clump
  properties are related to the disk, and show that their high masses
  and luminosities are a consequence of the high disk surface density.
\end{abstract}


\section{Introduction}

The majority of the stars in the most massive galaxies
(M$_{\star}\gsim$\,10$^{11}$\,M$_{\odot}$) formed around
8\,--\,10\,billion years ago, an epoch when star formation was at its
peak \citep{Hopkins06,Sobral12b}.  Galaxies at this epoch appear to be
gas-rich ($f_{\rm gas}$\,=\,20\,--\,80\%;
\citealt{Tacconi10,Daddi10,Geach11}) and turbulent \citep{Lehnert09},
with high velocity dispersions given their rotational velocities
($\sigma$\,=\,30\,--\,100\,km\,s$^{-1}$, $v_{\rm max}$\,/\,$\sigma\sim
$\,0.2\,--\,1;
e.g.\ \citealt{ForsterSchreiber09,Genzel08,Wisnioski11,Bothwell12}).
Within the dense and highly pressurised inter-stellar medium (ISM) of
these high-redshift galaxies, it has been suggestes that star formation
may be triggered by fragmentation of dynamically unstable gas (in
contrast to star-formation occurring in giant molecular clouds in the
Milky-Way which continually condense from a stable disk and then
dissipate).  This process may lead to the to the formation of massive
($\sim $\,10$^{8-9}$\,M$_{\odot}$) star-forming regions
\citep[e.g.\ ][]{ElmegreenD07,Bournaud09} and give rise to the the
clumpy morphologies that are often seen in high-redshift starbursts
\citep{Elmegreen09}.

In order to explain the ubiquity of ``clumpy'' disks seen in images of
high-redshift galaxies, numerical simulations have also suggested that
most massive, star-forming galaxies at $z$\,=\,1\,--\,3 continually
accrete gas from the inter-galactic medium along cold and clumpy
streams from the cosmic web
\citep{Keres05,Dekel09,Bournaud09,VandeVoort11}.  This mode of
accretion is at its most efficient at $z\sim $\,1\,--\,2, and offers a
natural route for maintaining the high gas surface densities, star
formation rates and clumpy morphologies of galaxies at these epochs.
In such models, the gas disks fragment into a few bound clumps which
are a factor 10\,--\,100\,$\times$ more massive than star-forming
complexes in local galaxies.  
The gravitational release of energy as the most massive clumps form,
torques between in-spiraling clumps and energy injection from star
formation are all likely to contribute to maintaining the high
turbulence velocity dispersion of the inter-stellar medium (ISM)
\citep[e.g.\ ][]{Bournaud09,Lehnert09,Genzel08,Genzel11}.

In order to refine or refute these models, the observational challenge
is now to quantitatively measure the internal properties of
high-redshift galaxies, such as their cold molecular gas mass and
surface density, disk scaling relations, chemical make up, and
distribution and intensity of star formation.  Indeed, constraining the
evolution of the star formation and gas scaling relations with
redshift, stellar mass and/or gas fraction are required in order to
understand star formation throughout the Universe.  In particular, such
observations are vital to determine if the prescriptions for star
formation which have been developed at $z$\,=\,0 can be applied to the
rapidly evolving ISM of gas-rich, high-redshift galaxies
\citep{Krumholz10,Hopkins12b}.


To gain a census of the dominant route by which galaxies assemble the
bulk of their stellar mass within a well selected sample of
high-redshift galaxies, we have conducted a wide field (several
degree-scale) near-infrared narrow-band survey (the High-Z Emission
Line Survey; HiZELS) which targets H$\alpha$ emitting galaxies in four
precise ($\Delta z$\,=\,0.03) redshift slices: $z$\,=\,0.40, 0.84, 1.47
and 2.23
\citep{Geach08,Sobral09,Sobral10,Sobral11,Sobral12a,Sobral12b}.  This
survey provides a large, star formation limited sample of identically
selected H$\alpha$ emitters with properties ``typical'' of galaxies
which will likely evolve into $\sim $\,L$_{\star}$ galaxies by
$z$\,=\,0, but seen at a time when they are assembling the bulk of
their stellar mass, and thus at a critical stage in their evolutionary
history.  Moreover, since HiZELS was carried out in the best-studied
extra-galactic survey fields, there is a wealth of multi-wavelength
data, including 16\,--\,36 medium and broad-band photometry (from
rest-frame UV\,--\,mid-infrared wavelengths allowing robust stellar
masses to be derived), \emph{Herschel} 250\,--\,500$\mu$m imaging
(allowing bolometric luminosities and star formation rates to be
derived) as well as high-resolution morphologies for a subset from the
\emph{Hubble Space Telescope} CANDELS and COSMOS ACS surveys.

In this paper, we present adaptive optics assisted integral field
spectroscopy of nine star-forming galaxies selected from HiZELS.  The
galaxies studied here have H$\alpha$-derived star formation rates of
1\,--\,27\,M$_\odot$\,yr$^{-1}$ and will likely evolve into $\sim
$\,L$^{\star}$ galaxies by $z$\,=\,0.  They are therefore
representative of the high-redshift star-forming population.  We use
the data to explore the scaling relations between the star formation
distribution intensity and gas dynamics within the ISM, as well as the
properties of the largest star-forming regions.  We adopt a cosmology
with $\Omega_{\Lambda}$\,=\,0.73, $\Omega_{m}$\,=\,0.27, and
H$_{0}$\,=\,72\,km\,s$^{-1}$\,Mpc$^{-1}$ in which 0.12$''$ corresponds
to a physical scale of 0.8\,kpc at $z$\,=\,1.47, the median redshift of
our survey.  All quoted magnitudes are on the AB system.  For all of
the star formation rates and stellar mass estimates, we use a
\citet{Chabrier03} initial mass function (IMF).

\section{Observations}
\setcounter{footnote}{0}

Details of the target selection, observations and data-reduction are
given in \citet{Swinbank12a}.  Briefly, we selected nine galaxies from
HiZELS with H$\alpha$ fluxes
0.7\,--\,1.6\,$\times$\,10$^{-16}$\,erg\,s$^{-1}$\,cm$^{-2}$ (star
formation rates\footnotemark of SFR$_{\rm
  H\alpha}$\,=\,1\,--\,27\,M$_{\odot}$\,yr$^{-1}$) which lie within
30$''$ of bright (R\,$<$\,15) stars.  We performed natural guide star
adaptive optics (AO) observations with the SINFONI IFU between 2009
September and 2011 April in $\sim $\,0.6$''$ seeing and photometric
conditions with exposure times between 3.6 to 13.4\,ks.  At the three
redshift slices of our targets, $z$\,=\,0.84[2], $z$\,=\,1.47[6] and
$z$\,=\,2.23[1], the H$\alpha$ emission line is redshifted to $\sim
$\,1.21, 1.61 and 2.12$\mu$m (i.e.\ into the $J$, $H$ and $K$-bands
respectively).  The median strehl achieved for our observations is 20\%
and the median encircled energy within 0.1$''$ (the approximate spatial
resolution of our observations) is 25\%.

\footnotetext{Adopting SFR$_{\rm
    Ha}$\,=\,4.6\,$\times$\,10$^{-42}$\,L(H$\alpha$) (erg\,s$^{-1}$)}

The data were reduced using the SINFONI {\sc esorex} data reduction
pipeline which extracts, flat-fields, wavelength calibrates and forms
the data-cube for each exposure.  The final (stacked) data-cube for each
galaxy was generated by aligning the individual data-cubes and then
combining them using an average with a 3-$\sigma$ clip to reject cosmic
rays.  For flux calibration, standard stars were observed each night
either immediately before or after the science exposures and were
reduced in an identical manner to the science observations.

\begin{figure*}
  \centerline{\psfig{file=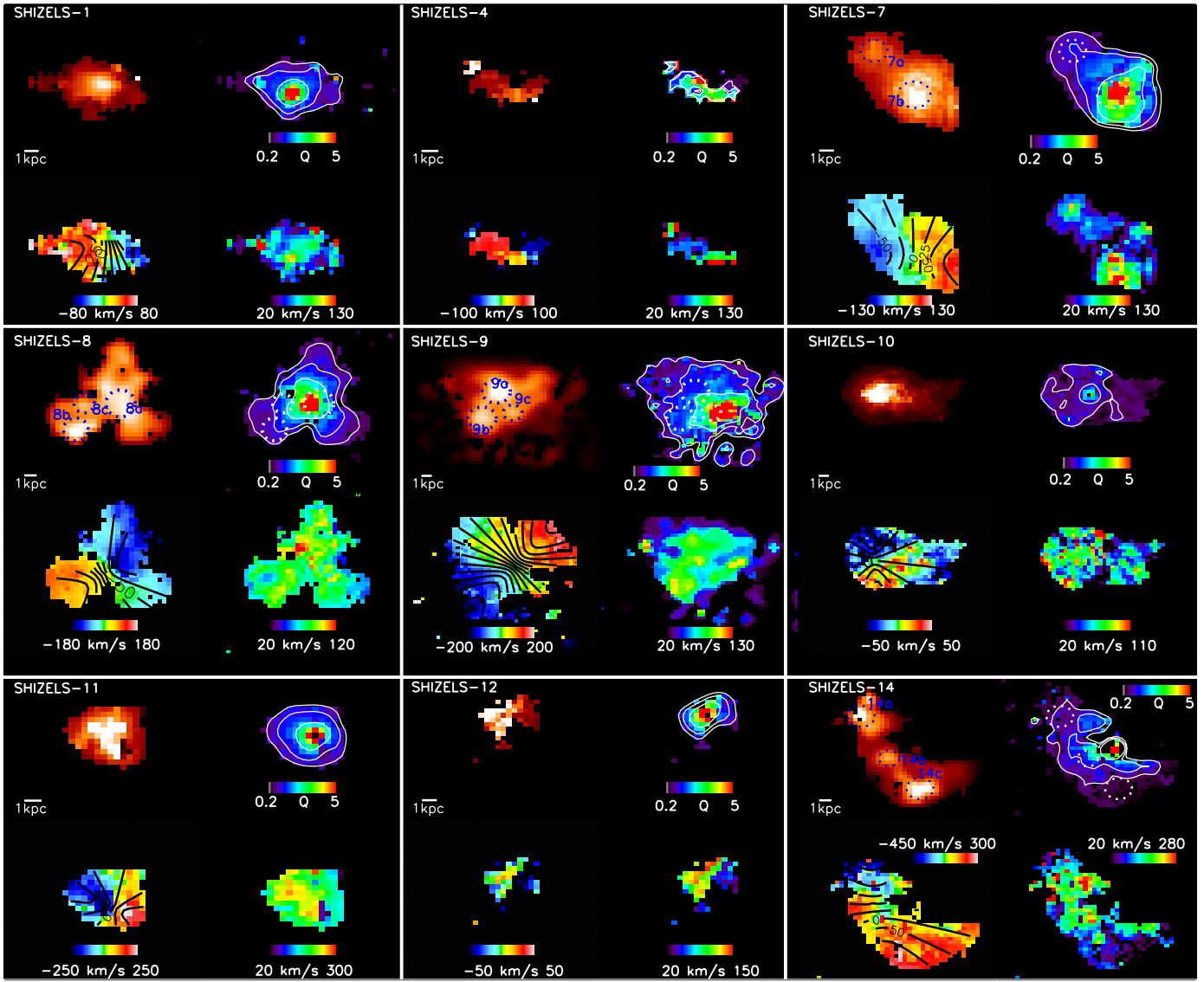,angle=90,width=8in}}
\caption{H$\alpha$ intensity, velocity field, line of sight velocity
  dispersion ($\sigma$) and Toomre ($Q$) maps of the nine SHiZELS
  galaxies in our sample.  {\it Top Left:} H$\alpha$ emission line map.
  In SHiZELS\,7, 8, 9, and 14 we identify and label the star-forming
  regions (clumps).  {\it Top Right:} Toomre $Q(x,y)$ maps of each
  galaxy, with contours at $Q$\,=\,0.5, 1.0, and 2.0.  In galaxies
  where we have identified star-forming regions (clumps), we also
  overlay their positions.  These star-forming regions have an average
  Toomre $Q$\,=\,0.8\,$\pm$\,0.4.  {\it Bottom Left:} H$\alpha$
  velocity field of each galaxy (with the best-fit kinematic model
  overlaid as contours).  {\it Bottom Right:} line of sight velocity
  dispersion ($\sigma$), corrected for local velocity gradient
  ($\Delta$\,V\,/\,$\Delta$\,R) across the PSF.  At least six galaxies
  (SHiZELS 1, 7, 8, 9, 10, \& 11), have dynamics that indicate that the
  ionised gas is in a large, rotating disk.  A further two are compact
  (SHiZELS 4 \& 12) whilst the dynamics of SHiZELS\,14 indicate a
  merger.}
\label{fig:2dmaps}
\end{figure*}

As Fig.~\ref{fig:2dmaps} shows, all nine galaxies in our SINFONI-HiZELS
survey (SHiZELS) display strong H$\alpha$ emission, with luminosities
of L$_{\rm H\alpha}\sim $\,10$^{41.4-42.4}$\,erg\,s$^{-1}$.  Fitting
the H$\alpha$ and [N{\sc ii}]$\lambda\lambda$6548,6583 emission lines
pixel-by-pixel using a $\chi^{2}$ minimisation procedure we construct
intensity, velocity and velocity dispersion maps of our sample and show
these in Fig.~\ref{fig:2dmaps} (see also \citealt{Swinbank12a} for
details).

\section{Analysis \& Discussion}

\subsection{Galaxy Dynamics and Star Formation}
\label{sec:dynSF}

As \citet{Swinbank12a} demonstrate, the ratio of
dynamical-to-dispersion support for this sample is
$v$\,sin($i$)\,/\,$\sigma$\,=\,0.3\,--\,3, with a median of
1.1\,$\pm$\,0.3, which is consistent with similar measurements for both
AO and non-AO studies of star-forming galaxies at this epoch
\citep[e.g.\ ][]{ForsterSchreiber09}.  The velocity fields and low
kinemetry values of the SHiZELS galaxies (total velocity asymmetry,
K$_{\rm tot}$\,=\,0.2\,--\,0.5) also suggest that at least six galaxies
(SHiZELS 1, 7, 8, 9, 10, \& 11) have dynamics consistent with large,
rotating disks, although all display small-scale deviations from the
best-fit dynamical model, with
$<$data\,$-$\,model$>$\,=\,30$\pm$10\,km\,s$^{-1}$, with a range from
$<$data\,$-$\,model$>$\,=\,15\,--\,70\,km\,s$^{-1}$
\citep{Swinbank12a}.

We also use the multi-wavelength imaging to calculate the rest-frame
SEDs of the galaxies in our sample and so derive the stellar mass,
reddenning and estimates of the star-formation history
\citep{Sobral11}.  From the broad-band SEDs (Fig.~1 of
\citealt{Swinbank12a}), the average E(B\,$-$\,V) for our sample is
E(B\,$-$\,V)\,=\,0.28\,$\pm$\,0.10 which corresponds to
A$_v$\,=\,1.11$\pm$0.27\,mag and indicates A$_{\rm
  H\alpha}$\,=\,0.91\,$\pm$\,0.21\,mag.  The resulting dust-corrected
H$\alpha$ star formation rate for the sample is SFR$_{\rm
  H\alpha}$\,=\,16\,$\pm$\,5\,M$_{\odot}$\,yr$^{-1}$, which is
consistent with that inferred from the far-infrared SEDs using stacked
{\it Herschel} SPIRE observations\footnotemark (SFR$_{\rm
  FIR}$\,=\,18\,$\pm$\,8\,M$_{\odot}$\,yr$^{-1}$;
\citealt{Swinbank12a})

\footnotetext{Adopting SFR$_{\rm
    FIR}$\,=\,2.7\,$\times$\,10$^{-44}$\,L$_{\rm FIR}$ (erg\,s$^{-1}$)}

\begin{figure*}
\centerline{\psfig{file=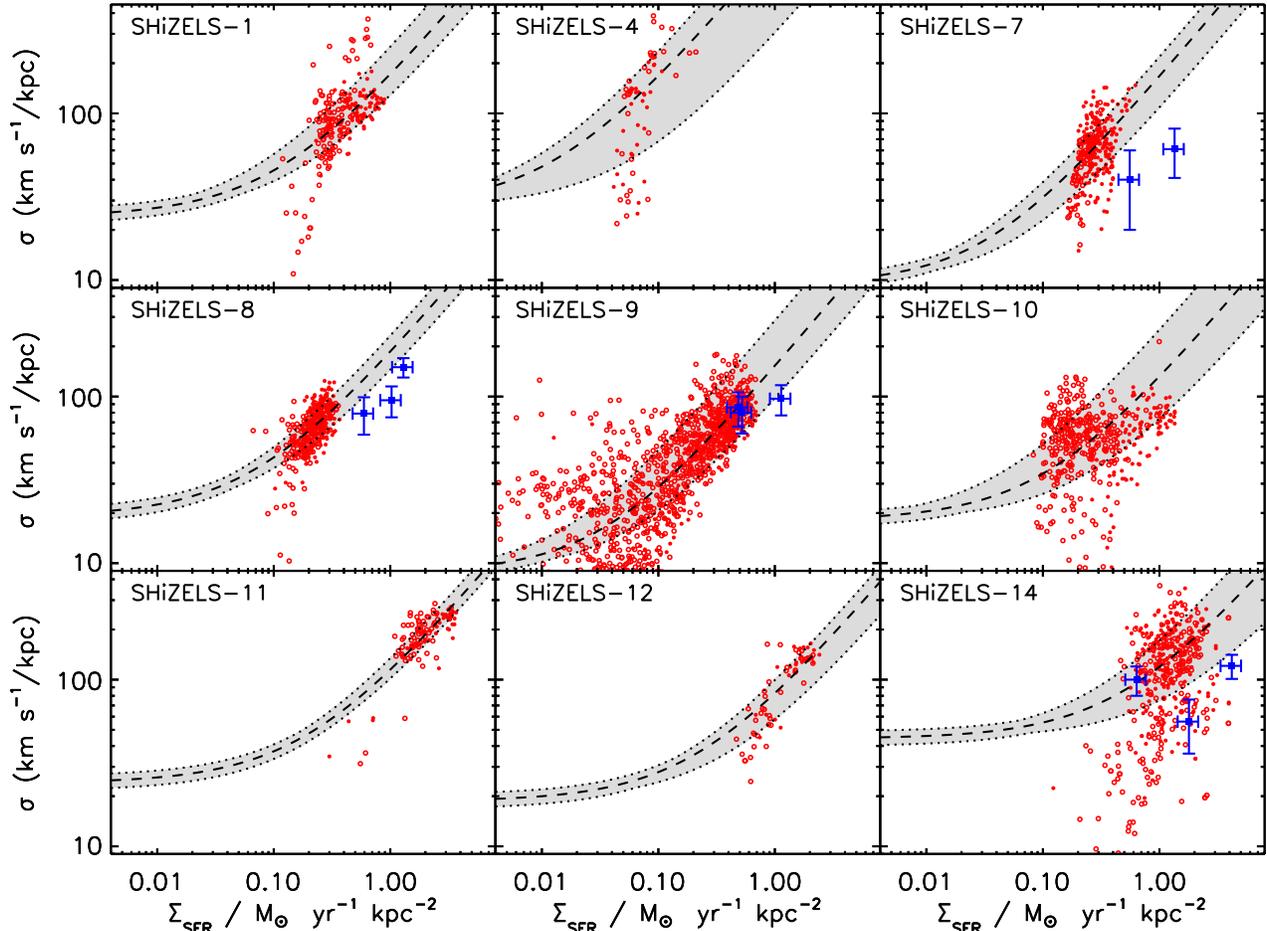,angle=90,width=6.5in}}
\caption{Star formation rate surface density as a function of velocity
  dispersion for each pixel within the galaxies in our sample.  The
  star formation rates are derived from H$\alpha$, corrected for galaxy
  reddening and the velocity dispersion has been corrected for local
  velocity gradient (\S~\ref{sec:dynSF}).  The small solid and open
  symbols denote measurements within and outside the half-light radius
  respectively.  The solid squares show the star-formation and velocity
  dispersions of the $\sim $\,kpc-scale clumps (Table~2) which appear
  as regions of high star formation density given their velocity
  dispersion.  The grey region denotes the best fit to the $\Sigma_{\rm
    SFR}$\,--\,$\sigma$ relation from combining the Toomre criterion
  and Kennicutt-Schmidt law (see equation~\ref{eqn:sS}) with power-law
  index ranging from $n$\,=\,1.0\,--\,1.4 (the dashed curve shows the
  solution for $n$\,=\,1.2).  Over this range, the data is consistent
  with an absolute star formation efficiency of
  $A$\,=\,4.1\,$\pm$\,2.4\,$\times$\,10$^{-4}$\,M$_{\odot}$\,yr$^{-1}$\,kpc$^{-2}$.}
\label{fig:SF_galgal}
\end{figure*}

Next, to investigate the star formation occurring within the ISM of
each galaxy, we measure the star formation surface density and velocity
dispersion of each pixel in the maps.  Since we do not have spatially
resolved reddening maps, for each galaxy we simply correct the star
formation rate in each pixel using the best-fit E(B\,$-$\,V) for that
system.  We also remove the rotational contribution to the line width
at each pixel by calculating the local $\Delta$V\,/\,$\Delta$R across
the point spread function (PSF) for each pixel \citep{Davies11}.  In
Fig.~\ref{fig:SF_galgal} we plot the resulting line of sight velocity
dispersion ($\sigma$) as a function of star formation surface density
($\Sigma_{\rm SFR}$) for each galaxy in our sample.  We see that there
appears to be a correlation between $\Sigma_{\rm SFR}$ and $\sigma$,
and as \citet{Krumholz10} show, this power-law correlation may be a
natural consequence of the gas and star formation surface density
scaling laws.  For example, first consider the Toomre stability
criterion, $Q$, \citep{Toomre64}.
\begin{equation}
  Q\,=\,\frac{\sigma\kappa}{\pi G\Sigma_{\rm disk}}
  \label{eqn:toomre}
\end{equation}
\noindent where $\sigma$ denotes the line of sight velocity dispersion,
$\Sigma_{\rm disk}$ is the average surface density of the disk,
$\kappa$\,=\,$a$\,$v_{\rm max}$\,/\,$R$ where $v_{\rm max}$ is the
rotational velocity of the disk, $R$ is the disk radius and
$a$\,=\,$\sqrt2$ for a flat rotation curve.  Galaxy whose disks have
$Q< $\,1 are unstable to local gravitational collapse and will fragment
into clumps, whereas those with $Q\gsim $\,1 have sufficient rotational
support for the gas to withstand and collapse.  As \citet{Hopkins12b}
\citep[e.g.\ see also][]{Cacciato12} point out, gas-rich galaxies are
usually driven to $Q\sim $\,1 since regions with $Q< $\,1 begin forming
stars, leading to super-linear feedback which eventually arrests
further collapse due to energy/momentum injection (recovering $Q\sim$
\,1).  For galaxies with $Q\gg $\,1, there is no collapse, no dense
regions form and hence no star formation (and so such galaxies would
not be selected as star-forming systems).

Following \citet{Rafikov01}, and focusing on the largest unstable
fluctuations, the appropriate combination of gas and stellar surface
density ($\Sigma_{\rm gas}$ and $\Sigma_{\star}$ respectively) is
\begin{equation}
  \Sigma_{\rm disk}\,=\,\Sigma_{\rm gas}\,+\,\left(\frac{2}{1+f_{\sigma}^2}\right)\Sigma_{\star}
   \label{eqn:KSlaw}
\end{equation}
where $f_{\sigma}$\,=\,$\sigma_{\star}$/$\sigma_{g}$ is the ratio of
the velocity dispersion of the stellar component to that of the gas
(see also the discussion in \citealt{Romeo11})..

Next, \citet{KS98} show that the gas and star formation surface
densities follow a scaling relation
\begin{equation}
  \left(\frac{\Sigma_{\rm SFR}}{\rm M_{\odot}\,yr^{-1}\,kpc^{-2}}\right)\,=\,A\left(\frac{\Sigma_{\rm gas}}{\rm M_{\odot}\,pc^{-2}}\right)^{\it n}
  \label{eqn:KS98}
\end{equation}
For local, star-forming galaxies, the exponent, $n\sim1.5$ and the
absolute star formation efficiency,
$A$\,=\,1.5\,$\pm$\,0.4\,$\times$\,10$^{-4}$ \citep{Kennicutt98} implying
an efficiency for star formation per unit mass of $\sim $\,0.04 which
holds across at least four orders of magnitude in gas surface density.

Combining these relations, the velocity dispersion, $\sigma$, should
therefore scale as
\begin{equation}
  \frac{\sigma}{\rm km\,s^{-1}}\,=\,\frac{\pi\,\times\,10^6\,G\,R}{\sqrt2\,v_{\rm max}}
  \left(\left(\frac{\Sigma_{\rm SFR}}{A}\right)^{1/n}+\left(\frac{2}{1+f_{\sigma^2}}\right)\frac{\Sigma_{\star}}{10^6}\,\right)
 \label{eqn:sS}
\end{equation}
\noindent where $\Sigma_{\rm SFR}$ and $\Sigma_{\star}$ are
measured in M$_{\odot}$\,yr$^{-1}$ and M$_{\odot}$\,kpc$^{-2}$
respectively, $R$ is in kpc, $v_{\rm max }$ in km\,s$^{-1}$, and
$G$\,=\,4.302\,$\times$\,10$^{-6}$\,kpc\,M$_{\odot}^{-1}$\,(km\,s$^{-1}$)$^2$.
With a power law index of $n$\,=\,1.4, and a marginally stable disk
($Q$\,=\,1), for each galaxy we therefore expect a power law relation
$\sigma\propto\Sigma_{\rm SFR}^{0.7}$\,+\,$constant$
\citep{Krumholz12}.

%
%
%
\begin{table*}
\begin{center}
{\footnotesize
{\centerline{\sc Table 1: Targets \& Galaxy Properties}}
\begin{tabular}{lcccccccccc}
\hline
\noalign{\smallskip}
ID          & RA           & Dec             & $z_{\rm H\alpha}$ &    SFR$_{\rm H\alpha}^{a}$ & $r_{1/2}^{b}$     & $\sigma_{\rm H\alpha}^{c}$ & $v_{\rm asym}^d$   &  E\,(B\,$-$\,V)   &  $\log$($\frac{M_{\star}}{M_{\odot}}$) & $\log$($\frac{M_{\rm gas}}{M_{\odot}}$) \\
            & (J2000)      & (J2000)         &                 &      (M$_{\odot}$/yr)     & (kpc)            & (km\,s$^{-1}$)        &  (km\,s$^{-1}$)  &                   &                                   & \\
            &              &                 &                 &                          &                  &                       &                 &                   &                                   & \\
\hline
SHiZELS-1   & 02\,18\,26.3 & $-$04\,47\,01.6 & 0.8425          &     2                    & 1.8\,$\pm$\,0.3  &  98$\pm$15            & 112$\pm$11      & 0.4\,$\pm$\,0.1   & 10.03\,$\pm$\,0.15                &  9.4\,$\pm$\,0.4 \\
SHiZELS-4   & 10\,01\,55.3 & $+$02\,14\,02.6 & 0.8317          &     1                    & 1.4\,$\pm$\,0.5  &  77$\pm$20            & ...             & 0.0\,$\pm$\,0.2   & 9.74\,$\pm$\,0.12                 &  8.9\,$\pm$\,0.4 \\
SHiZELS-7   & 02\,17\,00.4 & $-$05\,01\,50.8 & 1.4550          &     8                    & 3.7\,$\pm$\,0.2  &  75$\pm$11            & 145$\pm$10      & 0.2\,$\pm$\,0.2   & 9.81\,$\pm$\,0.28                 &  9.8\,$\pm$\,0.4 \\
SHiZELS-8   & 02\,18\,21.0 & $-$05\,19\,07.8 & 1.4608          &     7                    & 3.1\,$\pm$\,0.3  &  69$\pm$10            & 160$\pm$12      & 0.2\,$\pm$\,0.2   & 10.32\,$\pm$\,0.28                &  9.8\,$\pm$\,0.4 \\
SHiZELS-9   & 02\,17\,13.0 & $-$04\,54\,40.7 & 1.4625          &     6                    & 4.1\,$\pm$\,0.2  &  62$\pm$11            & 190$\pm$20      & 0.2\,$\pm$\,0.2   & 10.08\,$\pm$\,0.28                &  9.8\,$\pm$\,0.4 \\
SHiZELS-10  & 02\,17\,39.0 & $-$04\,44\,43.1 & 1.4471          &     10                   & 2.3\,$\pm$\,0.2  &  64$\pm$8             & 30$\pm$12       & 0.3\,$\pm$\,0.2   & 9.42\,$\pm$\,0.33                 &  9.9\,$\pm$\,0.4 \\
SHiZELS-11  & 02\,18\,21.2 & $-$05\,02\,48.9 & 1.4858          &     8                    & 1.3\,$\pm$\,0.4  &  190$\pm$18           & 224$\pm$15      & 0.5\,$\pm$\,0.2   & 11.01\,$\pm$\,0.24                & 10.1\,$\pm$\,0.4 \\
SHiZELS-12  & 02\,19\,01.4 & $-$04\,58\,14.6 & 1.4676          &     5                    & 0.9\,$\pm$\,0.5  &  115$\pm$10           & ...             & 0.3\,$\pm$\,0.2   & 10.59\,$\pm$\,0.30                &  9.6\,$\pm$\,0.4 \\
SHiZELS-14  & 10\,00\,51.6 &  +02:33\,34.5  & 2.2418           &     27                   & 4.6\,$\pm$\,0.4  &  131$\pm$17           & ...             & 0.4\,$\pm$\,0.1   & 10.90\,$\pm$\,0.20                & 10.1\,$\pm$\,0.4 \\     
\hline
Median      &    ...       &  ...           & 1.46             &     7\,$\pm$\,2          & 2.4\,$\pm$\,0.7  & 75$\pm$19             & 147$\pm$31      & 0.3\,$\pm$\,0.1   & 10.25\,$\pm$\,0.50                &  9.8\,$\pm$\,0.2 \\
\hline
\label{table:gal_props}
\end{tabular}
}
\end{center}
\vspace{-0.5cm} Notes: $^{a}$\,H$\alpha$ star formation rate using the
calibration from \citet{Kennicutt98} with a Chabrier IMF; SFR$_{\rm
  H\alpha}$\,=\,4.6\,$\times$\,10$^{-42}$\,L$_{\rm H\alpha}$.
$^b$\,H$\alpha$ half light radius, deconvolved for the PSF.
$^c$Average velocity dispersion for each galaxy, corrected for
beam-smearing due to the PSF. $^d$\,$v_{\rm asym}$ denotes the best-fit
asymptotic rotation speed of the galaxy, and is corrected for
inclination (see \citealt{Swinbank12a} for details on the kinematic
modeling of these galaxies).
\end{table*}

In order to test whether this model provides an adequate description of
our data, we fit the $\Sigma_{\rm SFR}$\,--\,$\sigma$ distribution for
each galaxy in our sample.  To estimate the stellar surface density,
$\Sigma_{\star}$, we we follow \citet{Sobral11} and perform a full SED
$\chi^2$ fit of the rest-frame UV\,--\,mid-infrared photometry using
the \citet{Bruzual03} and Bruzual (2007) population synthesis models.
We use photometry from up to 36 (COSMOS) and 16 (UDS) wide, medium and
narrow bands (spanning {\it GALEX} far-UV and near-UV bands to {\it
  Spitzer}/IRAC) and calculate the rest-frame spectral energy
distribution, reddening, star-formation history and stellar mass
\citep{Sobral10}.  The stellar masses of these galaxies range from
10$^{9.7-11.0}$\,M$_{\odot}$ (Table~1; see also \citealt{Swinbank12a}).

Since the stellar masses are calculated from 2$''$ aperture photometry
(and then corrected to total magnitudes using aperture corrections,
\citealt{Sobral10}), to estimate the stellar surface density in the
same area as our IFU observations, we assume that stellar light follows
an exponential profile with Sersic index, n$_{\rm serc}$\,=\,1\,--\,2
and calculate the fraction of the total stellar mass within the disk
radius, $R$ (which we define as two times the H$\alpha$ half light
radius, $r_{\rm h}$).  Allowing a range of power-law index from
$n$\,=\,1.0\,--\,1.8 and a ratio of stellar- to gas- velocity
dispersion of $f_{\sigma}$\,=\,1\,--\,2 \citep{Korchagin03}, we
calculate the best-fit absolute star formation efficiency, $A$ and in
Fig.~\ref{fig:SF_galgal} we overlay the best-fit solutions.  Over the
range $n$\,=\,1.0\,--\,1.8, the best fit absolute star formation
efficiency for the sample is
$A$\,=\,(4.1\,$\pm$\,2.4)\,$\times$10$^{-4}$\,M$_{\odot}$\,yr$^{-1}$\,kpc$^{-2}$
(where the error-bar incorporates the galaxy-to-galaxy variation, a
range of $f_{\sigma}$\,=\,1\,--\,2, and the errors on the stellar
masses of each galaxy).  We note that at low star formation rates and
stellar masses, there is a non-zero velocity dispersion due to the
sound speed ($c_s$) of the gas ($c_s\lsim $\,10\,km\,s$^{-1}$ for the
Milky Way at the solar circle) which we have neglected since this is
below both the resolution limit of our observations and the minimum
velocity dispersion caused the stellar disks in these systems.

\begin{figure*}
\centerline{\psfig{file=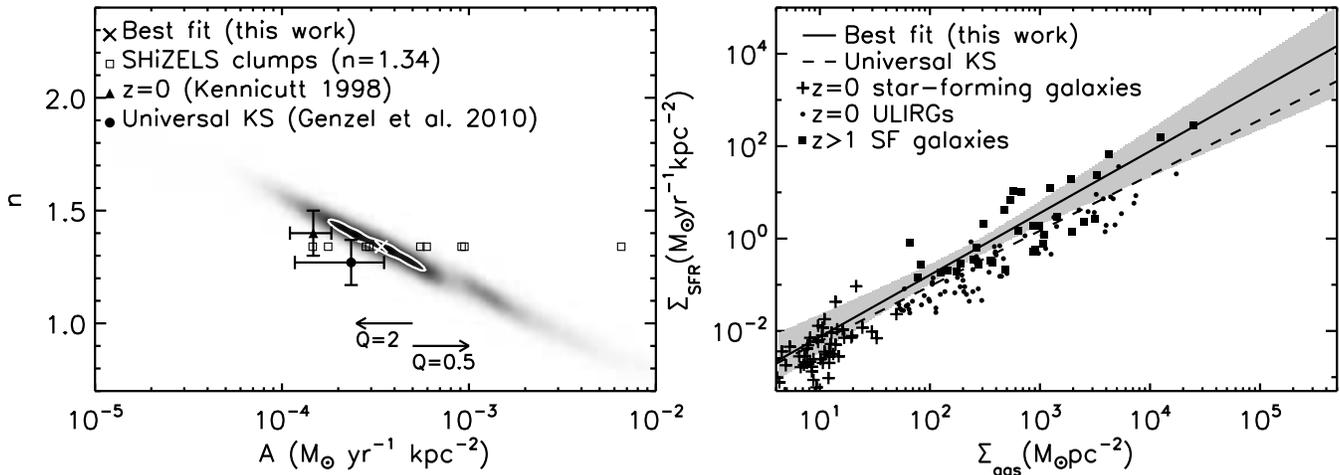,angle=90,width=7in}}
\caption{{\it Left:} The likelihood distribution for the power-law
  index ($n$) and absolute star formation efficiency ($A$) in the
  Kennicutt-Schmidt Law derived from the $\Sigma_{SFR}$\,--\,$\sigma$
  relations in Fig.~\ref{fig:SF_galgal} and assuming that the galaxies
  are marginally unstable, $Q$\,=\,1 (equation ~\ref{eqn:sS}) .  The
  best-fit solutions (within the 1\,$\sigma$ contour) have
  $n$\,=\,1.34\,$\pm$\,0.15 and
  $A$\,=\,3.4$_{-1.6}^{+2.5}$\,$\times$\,10$^{-4}$\,M$_{\odot}$\,yr$^{-1}$\,kpc$^{-2}$.
  The arrows shows how the absolute star formation efficiency would
  change if we adopt $Q$\,=\,0.5, or $Q$\,=\,2
  \citep[e.g.\ ][]{Leroy08}.  We also plot the position of the clumps
  (adopting $n$\,=\,1.34).  {\it Right:} The relation between star
  formation and gas-surface surface density for local- and
  high-redshift- star-forming galaxies and ULIRGs \citep{Genzel10}.
  The dashed line and shaded region shows the Kennicutt-Schmidt
  relation with our coefficients of $n$\,=\,1.34\,$\pm$\,0.15 and
  $A$\,=\,3.4$_{-1.6}^{+2.5}$\,$\times$\,10$^{-4}$\,M$_{\odot}$\,yr$^{-1}$\,kpc$^{-2}$.
  The solid line shows that best-fit solution for the ``Universal''
  relation from \citet{Genzel10}, which is well matched to our derived
  values.}
\label{fig:KS}
\end{figure*}

We can improve these constraints further assuming that star formation
in each galaxy behaves in a similar way.  We reiterate that this model
assumes the star formation is occurring in a marginally Toomre stable
disk, where the star formation follows the Kennicutt-Schmidt Law.  Over
a range
$A$\,=\,10$^{-5}$\,--\,10$^{-2}$\,(M$_{\odot}$\,yr$^{-1}$\,kpc$^{-2}$)
and $n$\,=\,0.8\,--\,2.5 we construct a likelihood distribution for all
nine galaxies and then convolve these to provide a composite likelihood
distribution, and show this in Fig.~\ref{fig:KS}.  Although the values
of $n$ and $A$ are clearly degenerate, the best-fit solutions have
$n$\,=\,1.34$\pm$0.15 and
$A$\,=\,3.4$_{-1.6}^{+2.5}$\,$\times$\,10$^{-4}$\,M$_{\odot}$\,yr$^{-1}$\,kpc$^{-2}$
Our derived values for the absolute star-formation efficiency, $A$, and
power-law index, $n$ are within the 1-$\sigma$ of the values derived
for local galaxies \citep[e.g.][]{KS98,Leroy08}.

Using the $^{12}$CO to trace the cold molecular gas, \citet{Genzel10}
showed that gas and star-formation surface densities of high-redshift
($z\sim $\,1.5) star-forming galaxies and ULIRGs are also well
described by the Kennicutt-Schmidt relation with coefficients
$n$\,=\,1.17\,$\pm$\,0.10 and
A\,=\,(3.3\,$\pm$\,1.5)\,$\times$10$^{-4}$\,M$_{\odot}$\,yr$^{-1}$\,kpc$^{-2}$,
which is comparable to the coefficients we derive from our sample.  

In Fig.~\ref{fig:KS} we plot the star formation and gas-surface surface
density for both local and high-redshift star-forming galaxies and
ULIRGs from \citet{Genzel10} and overlay the range of acceptable
solutions implied by our data.  We reiterate that we have adopted
$Q$\,=\,1 for this analysis and note that if we adopt $Q<$ \,1 then the
absolute star formation efficiency will be increased proportionally (as
shown in Fig.~\ref{fig:KS}).  Nevertheless, this shows that the values
of $n$ and $A$ we derive are consistent with the local and
high-redshift star-forming galaxies and ULIRGs, but free from
uncertainties associated with converting $^{12}$CO luminosities to
molecular gas mass, CO excitation or spatial extent of the gas
reservoir.

Using the values of $n$ and $A$ we have derived, we infer cold
molecular gas masses for the galaxies in our sample of M$_{\rm
  gas}$\,=\,10$^{9-10}$\,M$_{\odot}$ with a median M$_{\rm
  gas}$\,=\,7\,$\pm$\,2\,$\times$\,10$^{9}$\,M$_{\odot}$.  This
suggests a cold molecular gas fraction of M$_{\rm gas}$\,/\,(M$_{\rm
  gas}$\,+\,M$_{\star}$)\,=\,0.3\,$\pm$\,0.1 but with a range of
10\,--\,75\%, similar to those derived for other high-redshift
starbursts in other surveys \citep{Tacconi10,Daddi10,Swinbank11}.

\begin{figure}
  \centerline{\psfig{file=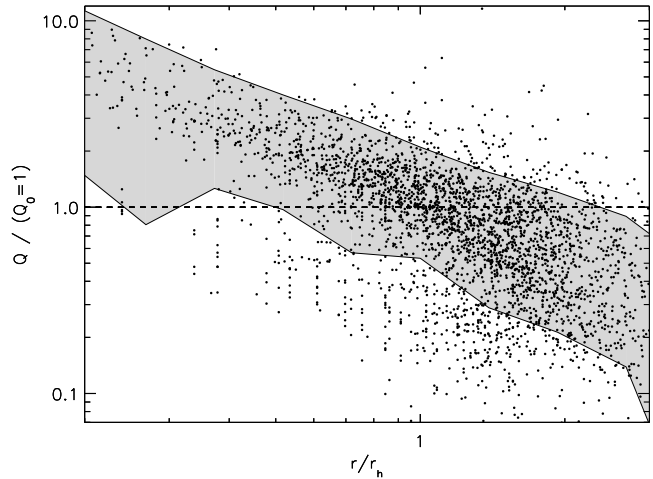,angle=90,width=3.5in}}
  \caption{The variation in Toomre parameter ($Q(x,y)$) within the ISM
    of the nine galaxies in our sample as a function of (normalised)
    radius.  The solid points denote the measurements at each pixel
    within each galaxy and the grey region shows the 18 and 81\,\%-ile
    limits of the distribution.  By construction, the average Toomre
    $Q$ in the sample is $Q(x,y)$\,=\,1, but varies by a factor $\sim
    $\,10 within the ISM, with the highest-$Q$ (most stable) in the
    central regions.}
\label{fig:Q_r}
\end{figure}

Finally, with estimates of the disk surface density, we can use
Eq.~\ref{eqn:toomre} to construct maps of the spatially resolved Toomre
parameter, $Q(x,y)$.  Since we set $Q$\,=\,1 to derive the coefficients
$n$ and $A$, by construction the average $Q$ across the population is
unity, but the relative range of $Q(x,y)$ within the ISM of each galaxy
is unaffected by this assumption.  In Fig.~\ref{fig:2dmaps} we show the
maps of $Q(x,y)$ for each galaxy in our sample (with contours marking
$Q(x,y)$\,=\,0.5, 1.0 and 2.0).  This shows that there is a range of
Toomre $Q$ across the ISM, and to highlight the variation with radius,
in Fig.~\ref{fig:Q_r} we show the Toomre parameter within each pixel of
each galaxy as a function of radius (normalised to the half light
radius, $r_{\rm h}$).  This shows that in the central regions, on
average the Toomre $Q$ increases by a factor $\sim $\,4\,$\times$
compared to $Q$ at the half light radius, whilst a radii greater than
$r_{\rm h}$, $Q$ decreases by approximately the same factor.

\subsubsection{Identification of Star-Forming Regions}
\label{sec:SFregions}

As Fig.~\ref{fig:2dmaps} shows, the galaxies in our sample exhibit a
range of H$\alpha$ morphologies, from compact (e.g.\ SHiZELS\,11 \& 12)
to very extended/clumpy (e.g.\ SHiZELS\,7, 8, 9 \& 14).  To identify
star-forming regions on $\sim $\,kpc scales and measure their basic
properties we isolate the star-forming clumps above the background
($\sigma_{\rm bg}$) by first converting the H$\alpha$ flux map into
photon counts (accounting for telescope efficiency) and then search for
3$\sigma_{\rm bg}$ over-densities above the radially averaged
background light distribution.  In this calculation, we demand that any
region is at least as large as the PSF.  We identify eleven such
regions and highlight these in Fig.~\ref{fig:2dmaps}.

It is still possible that selecting star-forming regions in this way
may give misleading results due to random associations and
signal-to-noise effects.  We therefore use the H$\alpha$ surface
brightness distribution from the galaxies and randomly generate $10^5$
mock images to test how many times a ``clump'' is identified.  We find
that only 2\,$\pm$\,1 spurious clumps (in our sample of eleven
galaxies) could be random associations.

Next, we extract the velocity dispersion and luminosity of each clump
from the using an isophote defining the star-forming region and report
their values in Table~2 (the clump velocity dispersions have been
corrected for the local velocity gradient from the galaxy dynamics and
sizes are deconvolved for the PSF).  Using the velocity dispersion and
star formation density of each clump, and fixing the power-law index in
the Kennicutt-Schmidt relation to $n$\,=\,1.34, we compute their
absolute star formation efficiencies, deriving a median $A_{\rm
  clump}$\,=\,5.4\,$\pm$\,1.5\,$\times$\,10$^{-4}$ (Fig.~\ref{fig:KS}).
This corresponds to an offset (at fixed $n$) from the galaxy-average of
$A_{\rm clump}$\,/\,$A$\,=\,1.3\,$\pm$\,0.4.  Equivalently, if we fix
the absolute star formation efficiency to that of the galaxy-average,
then the Toomre parameter in these regions is $Q$\,=\,0.8\,$\pm$\,0.4.

\begin{table}
{\footnotesize
{\centerline{\sc Table 2: Physical Properties of the Star-Forming Clumps}}
\smallskip
\begin{tabular}{lcccc}
\hline
\noalign{\smallskip}
Galaxy      & SFR                      & $\sigma_{\rm H\alpha}$ & [N{\sc ii}]/H$\alpha$  & $r_{\rm h}$ \\ 
            & (M$_{\odot}$\,yr$^{-1}$)   & (km\,s$^{-1}$)       &                        & (kpc) \\
\hline
SHiZELS-7    & 0.5\,$\pm$\,0.1          &  40\,$\pm$\,10   & 0.07\,$\pm$\,0.03         & 0.8\,$\pm$\,0.2 \\
SHiZELS-7    & 1.3\,$\pm$\,0.1          &  61\,$\pm$\,12   & 0.34\,$\pm$\,0.03         & 1.0\,$\pm$\,0.2 \\
SHiZELS-8    & 2.0\,$\pm$\,0.1          &  79\,$\pm$\,10   & 0.36\,$\pm$\,0.03         & 0.7\,$\pm$\,0.2 \\
SHiZELS-8    & 1.6\,$\pm$\,0.2          &  95\,$\pm$\,14   & 0.26\,$\pm$\,0.04         & 0.8\,$\pm$\,0.2 \\
SHiZELS-8    & 1.9\,$\pm$\,0.1          & 140\,$\pm$\,20   & 0.21\,$\pm$\,0.04         & 0.9\,$\pm$\,0.2 \\
SHiZELS-9    & 2.1\,$\pm$\,0.2          &  97\,$\pm$\,15   & 0.31\,$\pm$\,0.04         & 0.7\,$\pm$\,0.2 \\
SHiZELS-9    & 2.3\,$\pm$\,0.1          &  80\,$\pm$\,10   & 0.26\,$\pm$\,0.03         & 1.3\,$\pm$\,0.2 \\
SHiZELS-9    & 0.9\,$\pm$\,0.1          &  86\,$\pm$\,14   & 0.40\,$\pm$\,0.03         & $< $\,0.7 \\
SHiZELS-14   & 0.5\,$\pm$\,0.1          &  56\,$\pm$\,12   & 0.12\,$\pm$\,0.04         & 0.9\,$\pm$\,0.2 \\
SHiZELS-14   & 1.1\,$\pm$\,0.2          & 121\,$\pm$\,20   & 0.24\,$\pm$\,0.03         & $< $\,0.7 \\
SHiZELS-14   & 0.2\,$\pm$\,0.1          & 100\,$\pm$\,25   & $-$0.03\,$\pm$\,0.05         & 0.9\,$\pm$\,0.3 \\
\hline
Median       & 1.4\,$\pm$\,0.4          & 88\,$\pm$\,9     & 0.24\,$\pm$\,0.06          & 0.85\,$\pm$\,0.10 \\
\hline
\label{table:clumps}
\end{tabular}
}
\\
Notes: Half light radius, r$_h$, is deconvolved for PSF and the
velocity dispersion, $\sigma$, is corrected for local velocity gradient
(see \S~\ref{sec:dynSF}).  The star formation rates (SFR) are
calculated from the H$\alpha$ line luminosity using SFR$_{\rm
  H\alpha}$\,=\,4.6\,$\times$\,10$^{-42}$\,L$_{\rm H\alpha}$.
\end{table}

\subsection{The Scaling Relations of Local and High-Redshift Star-Forming Regions}
\label{sec:scaling}

The internal kinematics and luminosities of H{\sc ii} regions in local
galaxies, derived from the line widths of their emission lines, have
been the subject of various studies for some time
\citep[e.g.][]{Terlevich81,Arsenault90,Rozas98,Rozas06,Relano05}.  In
particular, if the large line widths of star-forming H{\sc ii} regions
reflect the virialization of the gas then they can be used to determine
their masses.  However, it is unlikely that this condition holds
exactly at any time during the evolution of a H{\sc ii} region due to
the input of radiative and mechanical energy, principally from their
ionizing stars \citep[e.g.\ ][]{Castor75}.  Nonetheless, the least
evolved H{\sc ii} regions may well be within a factor of a few (2\,--\,3)
of having their kinematics determined by their virial masses (at an
early stage, the stellar ionizing luminosities are maximized whereas
the mechanical energy input is minimized; \citealt{Leitherer99}).  In
the case of H{\sc ii} regions close to virial equilibrium, the use of
the line-width to compute gaseous masses offers a relatively direct
means to study the properties since it is independent of the
small-scale structure (density, filling factor, etc.).

\citet{Terlevich81} showed that the H$\beta$ luminosity of the most
luminous H{\sc ii} regions varies as
L(H$\beta$)\,$\propto\sigma^{4.0\pm0.8}$.  This result suggests that
the most luminous H{\sc ii} regions are likely to be virialized, so
that information about their masses, and the resultant mass-luminosity
relation, could be obtained using the virial theorem (they also claimed
a relation between a radius parameter and the square of the velocity
dispersion $\sigma$ for H{\sc ii} regions, as further evidence for
virialization).  However, more recent studies, in particular by
\citet{Rozas06} suggest that in super-giant H{\sc ii} regions,
L\,$\propto\sigma^{2.9\pm0.2}$ may be a more appropriate scaling (the
lower exponent arises since H{\sc ii} regions with the largest
luminosities are generally density-bound, which means that a
significant fraction of the ionizing radiation escapes and so does not
contribute to the luminosity, making shallower slopes physically
possible).

\begin{figure*}
  \centerline{\psfig{file=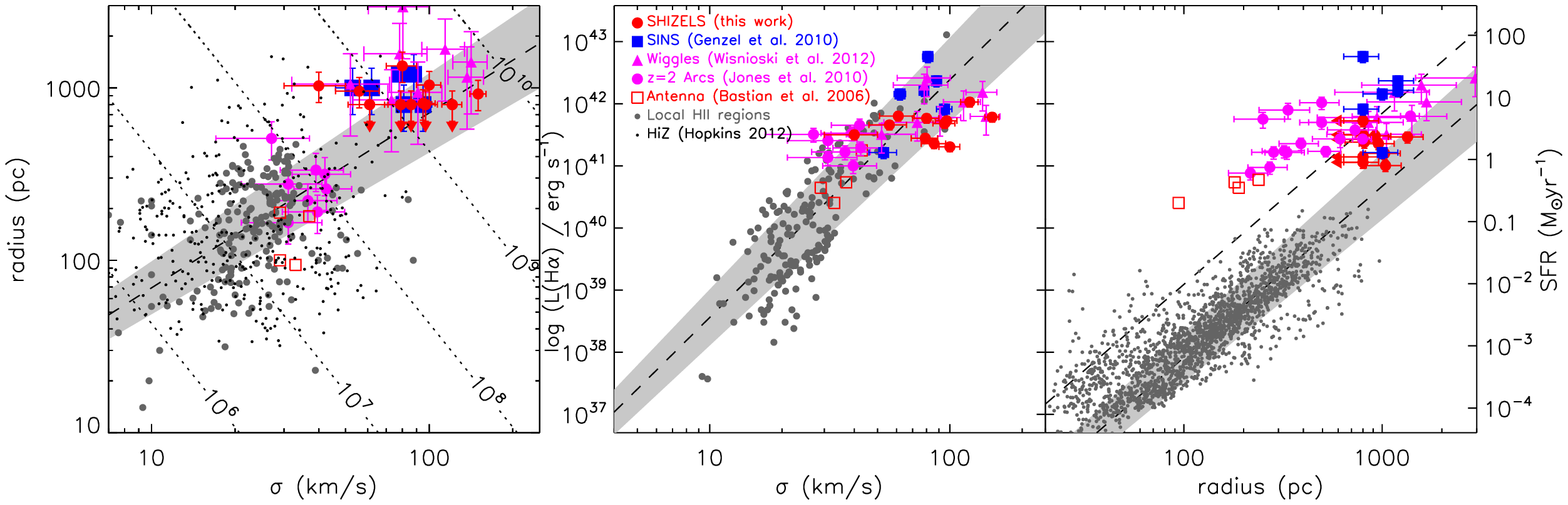,angle=90,width=7.0in}}
  \caption{Scaling relations between size, luminosity and velocity
    dispersion for the star-forming regions in our high-redshift
    galaxies compared to those in local GMCs and H{\sc ii} regions.  In
    all of these plots, we baseline our measurements against local data
    from
    \citet{Terlevich81,Arsenault90,Bordalo11,Fuentes-Masip00,Rozas06}.
          {\it Left:} The relation between velocity dispersion and
          size.  For the high-redshift star-forming regions, we also
          include clump measurements from SINS \citep{Genzel11},
          ZWiggles \citep{Wisnioski11b} and the cluster arc survey from
          \citep{Jones10}.  We also plot the properties of the HiZ GMCs
          from the numerical simulations from \citet{Hopkins12b}.  The
          dashed line shows a fit to the data of the form
          $r\propto\sigma^{1.01}$.  The dashed lines show lines of
          constant gas mass (~\ref{eqn:sigma_Sigma}).  {\it Middle:}
          The relation between velocity dispersion and luminosity of
          star-forming regions in high-redshift galaxies compared to
          those locally.  The dashed line denotes
          L$\propto\sigma^{3.8}$ which provides a good match to both
          the local and high-redshift data.  {\it Right:} The scaling
          relation between size and luminosity of star-forming regions.
          The high high-redshift star-forming regions have luminosity
          densities which are a factor $\sim $\,15\,$\pm$\,5\,$\times$
          higher than those typically found locally (see also
          \citealt{Wisnioski11b}).}
\label{fig:scaling}
\end{figure*}

To investigate the scaling relations of star-forming regions, in
Fig.~\ref{fig:scaling} we show the relations between luminosity, size
and velocity dispersion of the clumps in our sample compared to Giant
Molecular Clouds (GMCs) and H{\sc ii} regions in the Milky Way and
local galaxies
\citep{Terlevich81,Arsenault90,Bordalo11,Fuentes-Masip00,Rozas06}.  In
this plot, we also include the measurements of giant star-forming
regions from other high-redshift star-forming galaxies at $z\sim $\,1
from \citet{Wisnioski11b}, the $z\sim $\,1\,--\,2 galaxies from SINS
\citep{Genzel11}, and the clumps identified in strongly lensed $z\sim
$\,1.5\,--\,3 galaxies from \citet{Jones10} and \citet{Stark08}.

Despite the scatter, the radius\,--\,$\sigma$ and
$\sigma$\,--\,Luminosity relations of the high-redshift clumps
approximately follow the same scaling relations as those locally, but
extending up to $\sim $\,kpc scales.  Indeed, including all of the
data-points in the fits, we derive the scaling between size ($r$),
luminosity ($L$) and velocity dispersion ($\sigma$) of
\begin{equation}
  \log\left(\frac{r}{\rm kpc}\right)\,=\,(1.01\,\pm\,0.08)\,\log\left(\frac{\sigma}{\rm km\,s^{-1}}\right)\,+\,(0.8\,\pm\,0.1)
  \label{eqn:rs}
\end{equation}
and
\begin{equation}
  \log\left(\frac{L}{\rm erg\,s^{-1}}\right)\,=\,(3.81\,\pm\,0.29)\log\left(\frac{\sigma}{\rm km\,s^{-1}}\right)\,+\,(34.7\pm0.4)
  \label{eqn:Ls}
\end{equation}
Equation ~\ref{eqn:rs} suggests $\sigma\propto R$.  If the clouds are
self-gravitating clouds with $\sigma\propto R$, then the virial density
is constant.  The relation L\,$\propto\sigma^{3.81\pm0.29}$ is in
reasonable agreement with the early work from \citet{Terlevich81}, and
steeper than that found for super-giant H{\sc ii} regions in local
galaxies \citep{Rozas06}, although the large error bars (on both the
local and high-redshift data) preclude any firm conclusions.  Clearly a
larger sample is required to confirm this result and/or test whether
the scatter in the data is intrinsic.

If the star-forming regions we have identified are short lived, then
these scaling relations effectively reflect initial collapse conditions
of the clump as it formed, since a clump can not evolve far from those
initial conditions \citep[e.g.\ ][]{Ceverino10}.  In this case, the
relation between radius, velocity dispersion and gas mass should follow
$r$\,=\,$\sigma^2$\,/\,($\pi G \Sigma_{\rm disk}$) (see
\S~\ref{sec:Sigma_disk_Sigma_clump}).  In Fig.~\ref{fig:scaling} we
therefore overlay contours of constant gas mass in the r\,--\,$\sigma$
plane, which suggests that the {\it initial} gas masses for the clumps
is $M_{\rm
  gas}^{initial}$\,=\,2\,$\pm$\,1\,$\times$\,10$^9$\,M$_{\odot}$ a
factor $\sim $\,1000\,$\times$ more massive then the star-forming
complexes in local galaxies (e.g.\ see also
\citealt{Elmegreen09,Genzel11,Wisnioski11b}).  Assuming our gas mass
estimates from \S~\ref{sec:dynSF}, then these star-forming regions may
contain as much as $\sim$10\,--\,20\% of the cold molecular gas in the
disk.

Turning to the relation between size and luminosity of the star-forming
regions, it is evident from Fig.~\ref{fig:scaling} that the star
formation densities of the high-redshift clumps higher than those
locally.  Indeed, local star-forming regions follows a scaling relation
\begin{equation}
  \log\left(\frac{L}{\rm erg\,s^{-1}}\right)\,=\,(2.91\,\pm\,0.15)\log\left(\frac{r}{\rm kpc}\right)\,+\,(32.1\,\pm\,0.3)
  \label{eqn:Lr}
\end{equation}
We do not have sufficient number of objects or the dynamic range to
measure both the slope and zero-point of the size-luminosity relation
in the high-redshift clumps, and so instead we fix the slope of the
local relation (which is $L\propto r^{2.91\pm0.15}$) and fit for the
zero-point evolution and obtain
\begin{equation}
  \log\left(\frac{L}{\rm erg\,s^{-1}}\right)\,=\,(2.91\,\pm\,0.15)\log\left(\frac{r}{\rm kpc}\right)\,+\,(33.2\,\pm\,0.4)
  \label{eqn:Lr_hiz}
\end{equation}
This suggests that high-redshift star-forming regions have luminosities
at a fixed size that are on average a factor 15\,$\pm$\,5$\times$
larger than those locally (see also
\citealt{Swinbank09,Swinbank10Nature,Jones10,Wisnioski11b}). We note
that high luminosities at fixed size have been found in local
starbursts, such as in the Antennae \citep{Bastian06}, whilst offsets
of factors $\sim $\,50\,$\times$ have been inferred for star-forming
regions in high-redshift galaxies
\citep[e.g.\ ][]{Swinbank09,Jones10,Wisnioski11b}.  

%
%
%

\subsection{The Relation between the Disk and Clump Properties}
\label{sec:Sigma_disk_Sigma_clump}

It is possible to relate the properties of the clumps to the overall
properties of the disk \citep[e.g.\ ][]{Hopkins12a}.  For example, the
velocity dispersion of the fastest growing Jeans unstable mode which
can not be stabilised by rotation in a gas disk is given by
\begin{equation}
  \sigma_t(R)^2\,=\,\pi\,G\,\Sigma_{\rm disk}\,R
  \label{eqn:sigma_Sigma}
\end{equation}
\citep[e.g.\ ][]{Escala08,Elmegreen09b,Dekel09,Genzel11,Livermore12a}.
The critical density for collapse ($\rho_c$), on scale $R$ from a
turbulent ISM is given by
\begin{equation} 
\rho_c\,=\,\frac{3}{4\,\pi\,R^{3}}\,M_{\rm J}\,\simeq\,\frac{9}{8\,\pi\,R^2\,G}\,\sigma_t(R)^2 
\label{eqn:MJeans} 
\end{equation}
where $\sigma_t(R)$ is the line of sight velocity turbulent velocity
dispersion and $M_{\rm J}$ is the Jeans mass.  
The critical density for collapse therefore scales as
\begin{equation}
  \rho_c(R)\,=\,\frac{9}{8R}\,\Sigma_{\rm disk}
  \label{eqn:R_SigmaDisk}
\end{equation}
Assuming that the cloud contracts by a factor $\simeq $\,2.5 as it
collapses, the post-collapse surface density of the cloud is
\begin{equation}
  \Sigma_{\rm cloud}\simeq\,10\,\rho_c R\simeq \,10\,\Sigma_{\rm disk}
  \label{eqn:Sigmacloud_Sigmadisk}
\end{equation}
\citep[see also][]{Livermore12a}.  Thus, the surface density of the
collapsed cloud is independent of radius and proportional to the
surface density of the disk, with the normalisation set by the collapse
factor and under the assumption $Q$\,=\,1.  \citet{Hopkins12b} show
that this model provides an reasonable fit to giant molecular clouds in
the Milky-Way, and further, suggests that the surface density (and
hence surface brightness) of clouds should increase with the surface
density of the disk.

Using our estimates of the stellar and gas masses and spatial extent of
the galaxies in our sample, we derive disk surface densities of
$\Sigma_{\rm
  disk}$\,=\,1.1\,$\pm$\,0.4\,$\times$\,10$^{9}$\,M$_{\odot}$\,kpc$^{-2}$,
and hence expect the mass surface densities of the star-forming regions
that form to have mass surface densities of $\Sigma_{\rm clump}\sim
$\,10$^{10}$\,M$_{\odot}$\,kpc$^{-2}$.  It is instructive to compare this
to the average mass surface density of the clumps.  For example,
assuming that their velocity dispersions are virial and adopting
$M_{\rm clump}$\,=\,$C\sigma^2 r_h$\,/\,$G$, using the average velocity
dispersion and size of the clumps (Table~2), we derive an average clump
mass surface density of $\Sigma_{\rm
  clump}$\,=\,8\,$\pm$\,2\,$\times$\,10$^9$\,M$_{\odot}$\,kpc$^{-2}$ with
$C$\,=\,5 (appropriate for a uniform density sphere).  Although this
calculation should be considered crude as it is unclear whether the
velocity dispersions we measure are virial, it is encouraging that the
predicted surface mass densities of the clumps are similar to those
inferred from their velocity dispersions and sizes.

Finally, \citet{Hopkins12b} (see also \citealt{Escala08} and
\citealt{Livermore12a}) show that for a marginally stable disk of
finite thickness, density structures on scales greater than $h$ will
tend to be stabilised by rotation which leads to an exponential cut off
of the clump mass function above
\begin{equation}
  M_0\,\simeq\,\frac{4\pi}{3}\,\rho_c(h)\,h^3\,=\,\frac{3\,\pi\,G^2}{2}\frac{\Sigma_{\rm disk}^3}{\kappa^4}
  \label{eqn:M0_1}
\end{equation}
or
\begin{equation}
  \frac{M_0}{M_{\odot}}\,=\,8.6\,\times\,10^{3}\left(\frac{\Sigma_{\rm disk}}{\rm 10\,M_{\odot}\,pc^{-2}}\right)^3\,\left(\frac{\kappa}{\rm 100\,km\,s^{-1}\,kpc}\right)^{-4}
  \label{eqn:M0_2}
\end{equation}
This suggests that the most massive clumps that can form in a disk
(``the cut off mass'') depends strongly on the disk surface density --
increasing the disk surface density increases mass of the clumps that
are able to form \citep[e.g.\ ][]{Escala11}.  However, there is also a
competing (stabilising) factor from the epicyclic frequency such that a
fixed radius, higher circular velocities reduce the mass of the largest
clumps able to form.

Applying equation~\ref{eqn:M0_2} to the Milky-Way, with a cold
molecular gas fraction of 10\%, $f_{\sigma}$\,=\,2 \citep{Korchagin03},
the average surface density is $\Sigma_{\rm
  disk}$\,=\,35\,M$_{\odot}$\,pc$^{-2}$ and for
$\kappa$\,=\,220\,km\,s$^{-1}$/\,8\,kpc \citep{Feast97} the cut off
mass should be $M_0\sim$\,10$^7$\,M$_{\odot}$, in good agreement with
the characteristic mass of the largest galactic GMCs
\citep[e.g.\ ][]{Stark06}

How does the cut-off mass for our high-redshift sample compare to local
galaxies?  For $f_{\sigma}$\,=\,2, and using the scaling relations
derived in \S~\ref{sec:dynSF} to estimate the gas mass (Table~1),
($A$\,=\,3.4\,$\times$10$^{-4}$\,M$_{\odot}$\,yr$^{-1}$\,kpc$^{-2}$ and
$n$\,=\,1.34) we derive a range of cut off masses of
M$_{0}$\,=\,0.3\,--\,30\,$\times$\,10$^{9}$\,M$_{\odot}$ (with a median
and error of the sample of
$M_{0}$\,=\,9\,$\pm$\,5\,$\times$\,10$^{9}$\,M$_{\odot}$).  This is
similar to the mass inferred for the brightest star-forming regions
seen in high-resolution images of other high-redshift galaxies
\citep{Elmegreen89,ElmegreenD07,Elmegreen09,Bournaud09,ForsterSchreiber11,Genzel11,Wisnioski11b},
and a factor $\sim $\,1000\,$\times$ higher than the largest
characteristic mass of a star-forming region in the Milky-Way.

\begin{figure}
  \centerline{\psfig{file=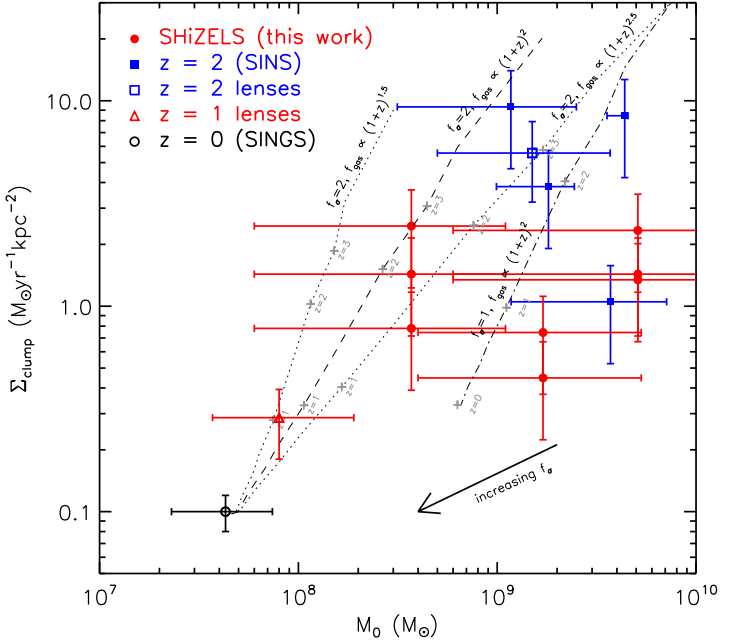,angle=90,width=3.8in}}
  \caption{The most massive clumps that can form (the ``cut off mass'',
    M$_{\rm 0}$) as a function of clump star formation surface density
    for SHiZELS galaxies.  The cut off mass is related to the disk
    surface density ($\Sigma_{\rm disk}$) and epicyclic frequency
    ($\kappa$) via M$_{\rm 0}\propto\Sigma_{\rm disk}^3\kappa^{-4}$.
    The $z$\,=\,0 observations are derived from the Spitzer Infrared
    Nearby Galaxy Survey (SINGS) \cite{Kennicutt03}.  We also include
    in the plot measurements of other high-redshift star-forming
    galaxies from the SINS survey \citep{Genzel11} and lensing surveys
    \citep{Jones10,Livermore12a}.  This shows that the
    cut off mass and star-formation surface densities of the
    high-redshift star-forming regions are (up to) a factor $\sim
    $\,100\,$\times$ higher than star-forming regions in local
    galaxies.  Using a simple model for galaxies with evolving gas
    fractions ($f_{\rm gas}\propto $\,(1\,+\,$z$)$^{(2\,\pm\,0.5)}$)
    and using the redshift evolution of disk scaling relations (size,
    rotational velocity and stellar mass) from \citet{Dutton11} and
    using equation ~\ref{eqn:Sigmacloud_Sigmadisk} and ~\ref{eqn:M0_2}
    can be used to derive model tracks to show how the cut off mass and
    clump star formation surface density are expected to evolve with
    redshift.  We plot these tracks for a ratio of stellar-to-gas
    velocity dispersion, $f_{\sigma}$\,=\,2, but also show how the
    results change if we instead adopted vary for $f_{\sigma}$\,=\,1.
    These tracks shows that the cut off mass and clump star formation
    surface density should increase by 1\,--\,2\,dex between $z$\,=\,0
    and $z$\,=\,3. }
\label{fig:mcut}
\end{figure}

In Fig.~\ref{fig:mcut} we plot our estimates of the the cut off mass
versus the clump star-formation densities for the galaxies in our
sample \citep[see also][]{Livermore12a}.  We use the H$\alpha$ derived
star-formation rate for each clump, corrected for galaxy reddening
(note that we do not have reddening estimates for individual clumps and
so we assume a factor 2\,$\times$ uncertainty in their star formation
surface density).  We also include estimates of the cut off mass and
star formation surface density from the SINS survey of $z\sim $\,2
galaxies from \citet{Genzel11} (with dynamics measured from
\citealt{ForsterSchreiber09} and \citealt{Cresci09}), as well as
measurements from the lensing samples of \citet{Livermore12a} ($z\sim
$\,1) and \citet{Jones10} ($z\sim $\,2).  Although the error-bars on
individual measurements are large (particularly due to the
uncertainties in deriving the gas surface density from the
Kennicutt-Schmidt law), as can be seen from Fig.~\ref{fig:mcut},
galaxies with high cut-off masses tend to have higher clump luminosity
surface densities.

It is also useful to adopt simple models for the evolution of galaxy
disks and gas fraction to investigate how the cut off mass and clump
properties may be expected to evolve with redshift.  For example,
\citet{Dutton11} present an analytic model for the evolution of disk
scaling relations (size, rotational velocity and stellar mass with
redshift; see \citealt{Dutton11} Table~3).  Combining with a simple
model for the evolution of the gas fraction $f_{\rm gas}\propto
$\,(1\,+\,$z$)$^{b}$ with $b$\,=\,1.5\,--\,2.5 \citep{Geach11} and
using Eq.~\ref{eqn:Sigmacloud_Sigmadisk} and Eq.~\ref{eqn:M0_2} we show
the expected evolution of the cut-off mass and clump luminosity surface
density with redshift.  This shows that as the gas fraction increases
(and adopting evolving models for the size, disk and circular velocity
of galaxies), then the cut off mass should increase by a factor
10\,--\,100,$\times$ over the redshift range $z$\,=\,0--2.5 whilst the
star formation density of the clumps should increase by a approximately
an order of magnitude over the same redshift range.  Although this is a
simple model, this framework allows us to understand why the properties
of the star-forming clumps within the ISM of our sample of
high-redshift galaxies are different to those typically found in
star-forming galaxies locally.

\section{Conclusions}

We have presented resolved spectroscopy of nine star-forming galaxies
at $z$\,=\,0.84\,--\,2.23 selected from the UKIRT/HiZELS survey.  These
galaxies have reddenning corrected star-formation rates of
SFR\,=\,16\,$\pm$5\,M$_{\odot}$\,yr$^{-1}$ and so are representative of
the high-redshift population \citep{Sobral12b}.  The H$\alpha$ dynamics
suggest that the ionised gas in at least six galaxies is in the form of
large, rotating disks.  We use the inferred rotation speeds of these
systems, together with the spatial extent of the H$\alpha$ to
investigate the star formation within the ISM, and we derive the
following main conclusions:

~$\bullet$ The star formation and velocity dispersion within the ISM of
these high-redshift galaxies follow a power-law relation of the form
$\sigma\propto A\Sigma_{\rm gas}^{1/n}$\,+\,$constant$ where the
coefficients, $A$ and $n$ are set by the Kennicutt-Schmidt Law
($\Sigma_{\rm SFR}$\,=\,$A\Sigma_{\rm gas}^n$) and the constant
includes the disk stellar surface density of finite thickness.
Assuming the gas disks are marginally stable ($Q$\,=\,1) we combine the
solutions for each galaxy and derive best-fit parameters of
power-law exponent, $n$\,=\,1.34$\pm$0.15 and absolute star formation
efficiency,
$A$\,=\,3.4$_{-1.6}^{+2.5}$\,$\times$\,10$^{-4}$\,M$_{\odot}$\,yr$^{-1}$\,kpc$^{-2}$.
These values are consistent with the parameters derived via $^{12}$CO
observations for both local and high-redshift star-forming galaxies,
but free from any assumptions about $^{12}$CO\,--\,H$_2$ conversion
factors, $^{12}$CO excitation or the spatial extent of the gas
reservoir.

~$\bullet$ Applying these coefficients, we infer cold molecular gas
masses in the range M$_{\rm gas}$\,=\,10$^{9-10}$\,M$_{\odot}$ with a
median M$_{\rm gas}$\,=\,7\,$\pm$\,2\,$\times$\,10$^{9}$\,M$_{\odot}$
and hence a cold molecular gas fraction of M$_{\rm gas}$\,/\,(M$_{\rm
  gas}$\,+\,M$_{\star}$)\,=\,0.3\,$\pm$\,0.1 but with a range of
10\,--\,75\%.

~$\bullet$ Using a simple analytic model, we show that the largest
structures that can form within the disk (the cut-off mass, $M_0$) are
set by the disk surface density with a competing (stabilising) force
from the epicyclic frequency such that M$_0\propto\Sigma_{\rm
  disk}^3\kappa^{-4}$.  For the galaxies in our sample, we derive cut
off masses of $M_0\sim $\,10$^{9}$\,M$_{\odot}$, a factor $\sim
$\,1000\,$\times$ higher than the largest characteristic mass of GMCs
in the Milky-Way.

~$\bullet$ Within the ISM of these galaxies, we reliably isolate eleven
$\sim $\,kpc-scale star-forming regions and measure their properites.
We show that their luminosities and velocity dispersions follow the
same scaling relations between size and velocity dispersion as local
H{\sc ii} regions.  Assuming the line widths are virial, the masses
derived for these star-forming regions are consistent with those
implied by the cut-off mass.  However, we find that the luminosity
densities of these star-forming regions are a factor $\sim
$\,15$\times$ higher than those typically found locally, which we
attribute to the requirement that the surface density of the
(collapsed) cloud must be $\sim$\,10\,$\times$ that of the disk.

Overall, the scaling relations we have derived suggest that the star
formation processes in high-redshift disks are similar to those in
local spiral galaxies, but occurring in systems with a gas rich and
turbulent ISM.  Given the paucity of gas-rich, clumpy disk-like
high-redshift galaxies
\citep{ElmegreenD07,Elmegreen09,ForsterSchreiber11b}, the next step in
these studies is to spatially resolve the cold molecular gas via CO
spectroscopy in a well selected sample in order to better constrain the
interation between star-formation and gas dynamics.  Through
comparisons with cosmologically based numerical simulations
\citep[e.g.\ ][]{Crain09,VandeVoort11,Ceverino10}, as well as high
resolution simulations of individual gas-rich disks
\citep[e.g.\ ][]{Agertz09,Krumholz10b} such observations may begin to
differentiate whether the dominant mode of accretion is via
three-dimensional cold gas flows accrete from the inter-galactic
medium, or from two-dimensions from outskirts of the disk as gas cools
from the hot halo.

\section*{acknowledgments}
We would like to thank the anonymous referee for their constructive
report which significantly improved the content and clarity of this
paper.  We thank Mario van der Ancker for help and support with the
SINFONI planning/observations, and Richard Bower, Avashi Dekel,
Reinhard Genzel, Rachael Livermore, Natascha F\"orster-Schreiber, and
Phil Hopkins for a number of very useful discussions.  AMS gratefully
acknowledges an STFC Advanced Fellowship.  DS is supported by a NOVA
fellowship.  IRS acknowledges support from STFC and a Leverhume Senior
Fellowship.  JEG is supported by a Banting Fellowship, administered by
the Natural Sciences and Engineering Research Council of Canada.  This
research was also supported in part by the National Science Foundation
under Grant No. NSF PHY11-25915.  The data presented here are based on
observations with the SINFONI spectrograph on the ESO/VLT under program
084.B-0300.


\end{document}